\documentclass[11pt]{article}
\newtheorem{example}{Example}[section]
\newtheorem{lemma}{Lemma}[section]
\newtheorem{note}{Note}[section]
\newtheorem{theorem}{Theorem}[section]
\newtheorem{corollary}{Corollary}[section]
\parskip=\baselineskip

\usepackage{amssymb}
\usepackage{amsmath}

\begin{document}
\begin{center}
\begin{Large}
{\LARGE\bf Expansions for Quantiles and Multivariate Moments}\\[1ex]
{\LARGE\bf of Extremes for}\\[1ex]
{\LARGE\bf Distributions of Pareto Type}\\[1ex]
\end{Large}
by\\[1ex]
Christopher S. Withers\\
Applied Mathematics Group\\
Industrial Research Limited\\
Lower Hutt, NEW ZEALAND\\[2ex]
Saralees Nadarajah\\
School of Mathematics\\
University of Manchester\\
Manchester M13 9PL, UK
\end{center}
\vspace{1.5cm}
{\bf Abstract:}~~Let $X_{nr}$  be the $r$th largest of a random sample of size $n$  from a
distribution $F (x) = 1 - \sum_{i = 0}^\infty c_i x^{-\alpha - i \beta}$ for $\alpha > 0$ and $\beta > 0$.
An inversion theorem is proved and used to derive an expansion for the quantile $F^{-1} (u)$ and powers of it.
From this an expansion in powers of $(n^{-1}, n^{-\beta/\alpha})$ is given for the multivariate moments of the extremes
$\{ X_{n, n - s_i}, 1 \leq i \leq k \}/n^{1/\alpha}$ for fixed ${\bf s} = (s_1, \ldots, s_k)$, where $k \geq 1$.
Examples include the Cauchy, Student $t$, $F$, second extreme distributions and stable laws of index $\alpha < 1$.

\noindent
{\bf AMS 2000 Subject Classification:}~~Primary 62E15; Secondary 62E17.

\noindent
{\bf Keywords and Phrases:}~~Bell polynomials; Extremes; Inversion theorem; Moments; Pareto; Quantiles.

\section{Introduction and Summary}

For $ 1 \leq r \leq n $, let $X_{nr}$ be the $r$th largest of a random
sample of size $n$ from a continuous distribution $F$  on $R$, the real numbers.
Let $f$  denote the density of $F$ when it exists.
The study of the asymptotics of the moments of $X_{nr}$ has been of considerable interest.
McCord (1964) gave a first approximation to the moments of $X_{n1}$ for three classes.
This showed that a moment of $X_{n1}$ can
behave like any positive power of $n$  or $ n_1 = \log \ n$.
(Here $\log$ is to the base $e$.)
Pickands (1968) explored the conditions under which various
moments of $(X_{n1} - b_n)/a_n$ converge to the corresponding moments of the extreme value distribution.
It was proved that this is indeed true for all $F$ in the domain of attraction of an
extreme value distribution provided that the moments are finite for sufficiently large $n$.
For other work, we refer the readers to Polfeldt (1970), Ramachandran (1984) and Resnick (1987).

The asymptotics of the quantiles of $X_{nr}$ have also been studied.
Note that $U_{nr} = F (X_{nr})$ is the $r$th order statistics from $U (0, 1)$.
For $1 \leq r_1 < r_2 < \cdots < r_k \leq n$ set $U_{n, {\bf r}} = \{ U_{n r_i}, 1 \leq i \leq k \}$.
By Section 14.2 of Stuart and Ord (1987), $U_n$ has the multivariate beta density
\begin{eqnarray}
U_{n, {\bf r}} \sim B ({\bf u} : {\bf r}) =
\prod_{i = 0}^k \left( u_{i + 1} - u_i \right)^{r_{i + 1} - r_i - 1}/B_n ({\bf r})
\label{1.1}
\end{eqnarray}
on $0 < u_1 < \cdots < u_k < 1$, where $u_0 = 0$, $u_{k + 1} = 1$, $r_0 = 0$, $r_{k + 1} = n + 1$ and
\begin{eqnarray}
B_n ({\bf r}) = \prod_{i = 1}^k B \left( r_i, r_{i + 1} - r_i \right).
\label{1.2}
\end{eqnarray}
David and Johnson (1954) expanded $X_{n r_i} = F^{-1} (U_{n r_i})$ about $u_{n i} = E U_{n r_i} = r_i/(n + 1)$:
$X_{n r_i} = \sum_{j = 0}^\infty G^{(j)} (u_{ni}) (U_{ni} - u_{ni})^j/j!$, where $G (u) = F^{-1} (u)$,
and using the properties of (\ref{1.1}) showed that if ${\bf r}$ depends on $n$ in such a ways that ${\bf r}/n \rightarrow {\bf p} \in ({\bf 0}, {\bf 1})$
as $n \rightarrow \infty$ then the $m$th order cumulants of $X_{n, {\bf r}} = \{ X_{n r_i}, 1 \leq i \leq k \}$
have magnitude $O (n^{1 - m})$ -- at least for $n \leq 4$, so that the distribution of $X_{n, {\bf r}}$ has a multivariate
Edgeworth expansion in powers of $n^{-1/2}$.
(Alternatively one can use James and Mayne (1962) to derive the cumulants of $X_{n, {\bf r}}$ from those of $U_{n, {\bf r}}$.)
The method requires the derivatives of $F$ at $\{ F^{-1} (p_i), 1 \leq i \leq k \}$ so breaks down if $p_i = 0$ or $p_k = 1$ --
which is the situation we study here.
For definiteness, we confine ourselves to $F^{-1} (u)$
having a power singularity at 1, say $F^{-1} (u) \sim (1 - u)^{-1/\alpha}$
as $u \to 1$, where $\alpha > 0$ that is,
\begin{eqnarray}
1 - F (x) \sim x^{-\alpha}
\label{1.3}
\end{eqnarray}
as $x \to \infty$.
For a nonparametric estimate of $\alpha$ see Novak and Utev (1990).

Distributions satisfying (\ref{1.3}) are known as Pareto type distributions.
These distributions arise in many areas of the sciences, engineering and medicine.
Some of these areas -- where publications involving Pareto type distributions have appeared -- are:
hydrology, physics, wind engineering and industrial aerodynamics, computer science,
water resources, insurance mathematics and economics, structural safety,
material science, performance evaluation, queueing systems, geophysical research,
ironmaking and steelmaking, banking and finance, atmospheric environment, civil engineering,
communications, information processing and management, high speed networks,
lightwave technology, solar energy engineering, supercomputing,
natural hazards and earth system sciences, ocean engineering, optics communications,
reliability engineering, signal processing and urban studies.

In Withers and Nadarajah (2007a) we showed that for fixed ${\bf r}$ when (\ref{1.3})
holds the distribution of $X_{n, n {\bf 1} - {\bf r}}$ (where ${\bf 1}$ is the vector of ones in $\Re^k$),
suitably normalized tends to a certain multivariate extreme value distribution as $n \rightarrow \infty$,
and so obtained the leading terms of the expansions of its moments in inverse powers of $n$.
Here we show how to extend those expansions when
\begin{eqnarray}
F^{-1} (u) = \sum_{i = 0}^\infty b_i (1 - u)^{\alpha_i}
\label{1.4}
\end{eqnarray}
with $\alpha_0 < \alpha_1 < \cdots$, that is,
$\{ 1 - F (x) \} x^{-1/\alpha_0}$ has a power series in
$\{ x^{-\delta_i} : \delta_i = (\alpha_i - \alpha_0)/\alpha_0 \}$.
Hall (1978) considered (\ref{1.4}) with $\alpha_i = i - 1/\alpha$,
but did not give the corresponding expansion for $F (x)$ or expansions in inverse powers of $n$.
He applied it to the Cauchy.
In Section 2, we demonstrate the method when
\begin{eqnarray}
1 - F (x) = x^{-\alpha} \sum_{i = 0}^\infty c_i x^{-i \beta},
\label{1.5}
\end{eqnarray}
where $\alpha > 0$ and $\beta > 0$.
In this case, (\ref{1.4}) holds with $\alpha_i = (i \beta - 1)/\alpha$.
In Section 3, we apply it to the Student $t$, $F$ and second extreme value distribution
and to stable laws of exponent $\alpha < 1$.
Appendix A gives the inverse theorem needed to pass from (\ref{1.5}) to (\ref{1.4}),
and expansions for powers and logs of series.

We use the following notation and terminology.
Let $(x)_i = \Gamma (x + i)/\Gamma (x)$ and $<x>_i = \Gamma (x + 1)/\Gamma (x - i + 1)$.
An inequality in $\Re^k$ consists of $k$ inequalities.
For example, for ${\bf x}$ in $C^k$, where $C$ is the set of complex numbers,
$Re ({\bf x}) < {\bf 0}$ means that $Re (x_i) < 0$ for $1 \leq i \leq k$.
Also $I (A) = 1$ or $0$ for $A$ true or false and $\delta_{ij} = I (i = j)$.
For $\boldsymbol{\theta} \in C^k$ let $\bar{\boldsymbol{\theta}}$ denote the vector with $\bar{\theta}_i = \sum_{j = 1}^k \theta_j$.

\section{Main Results}

For $1 \leq r_1 < \cdots < r_k \leq n$ set $s_i = n - r_i$.
Here, we show how to obtain expansions in inverse powers of $n$ for the
moments of the $X_{n, {\bf s}}$ for fixed ${\bf r}$ when (\ref{1.4}) holds,
and in particular when the upper tail of $F$ satisfies (\ref{1.5}).

\begin{theorem}
Suppose (\ref{1.5}) holds with $c_0$, $\alpha$, $\beta > 0$.
Then $F^{-1} (u)$ is given by (\ref{1.4}) with $\alpha_i = i a - 1/\alpha$,
$a = \beta/\alpha$ and $b_i = C_{i, 1/\alpha}$, where
$C_{i \psi} = c_0^\psi \widehat{C}_i (-\psi, c_0, x^{*}$ of (\ref{A4}) and
$x_i^{*} = x_i^{*} (a, 1, c)$ of (\ref{A15}):
\begin{eqnarray*}
C_{0 \psi}
&=&
c_0^\psi,
\\
C_{1 \psi}
&=&
\psi c_0^{\psi - a - 1} c_1,
\\
C_{2 \psi}
&=&
\psi c_0^{\psi - 2 a - 2} \left\{ c_0 c_2 + (\psi - 2 a - 1) c_1^2/2 \right\},
\\
C_{3 \psi}
&=&
\psi c_0^{\psi - 3 a - 3} \left[ c_0^2 c_2 + (\psi - 3 a - 1) c_0 c_1 c_2 + \left\{ (\psi + 1)_2/6 (\psi + 3a/2) (a+1) \right\} c_1^3 \right],
\end{eqnarray*}
and so on.
Also for any $\theta$ in $\Re$,
\begin{eqnarray}
\left\{ F^{-1} (u) \right\}^\theta =
\sum_{i = 0}^\infty (1 - u)^{i a - \psi} C_{i \psi}
\label{2.1}
\end{eqnarray}
at $\psi = \theta/\alpha$.
\end{theorem}

\begin{note}
On those rate occasions where the coefficients $d_i = C_{i, 1/\alpha}$ in
$F^{-1} (u) = \sum_{i = 0}^\infty (1 - u)^{i a - 1/\alpha} d_i$ are known from some alternative
formula then one can use $C_{i \psi} = d_0^\theta \widehat{C}_i (\theta, 1/d_0, d)$ of (\ref{A4}).
\end{note}

\noindent
{\bf Proof of Theorem 2.1}
By Theorem A.1 with $k = 1$, $u = x^{-\alpha}$, $x = c$,
we have $x^{-\alpha} = \sum_{i = 0}^\infty x_i^{*} (1 - u)^{1 + i a}$ at $u = F (x)$, where
\begin{eqnarray*}
x_0^{*}
&=&
c_0^{-1},
\\
x_1^{*}
&=&
c_0^{-a - 2} c_1,
\\
x_2^{*}
&=&
c_0^{-2 a - 3} \left\{ -c_0 c_2 + (a + 1) c_1^2 \right\},
\\
x_3^{*}
&=&
c_0^{-3 a - 4} \left\{ -c_0^2 c_3 + (2 + 3 a) c_0 c_1 c_2 - (2 + 3a) (1 + a) c_1^2/2 \right\},
\end{eqnarray*}
and so on.
So, for $S$ of (\ref{A1}), $x^{-\alpha} = c_0^{-1} v (1 + c_0 S (v^a, x^{*}))$ at $v = 1 - u$.
Now apply (\ref{A3}).
$\ \Box$

\begin{lemma}
For $\boldsymbol{\theta}$ in $C^k$,
\begin{eqnarray}
E \prod_{i = 1}^k \left( 1 - U_{n, r_i} \right)^{\theta_i} = b_n \left( {\bf r} : \bar{\boldsymbol{\theta}} \right),
\label{2.2}
\end{eqnarray}
where
\begin{eqnarray}
b_n \left( {\bf r} : \bar{\boldsymbol{\theta}} \right) =
\prod_{i = 1}^k b \left( r_i - r_{i - 1}, n - r_i + 1 : \bar{\theta}_i \right)
\label{2.3}
\end{eqnarray}
and $b (\alpha, \beta : \theta = B (\alpha, \beta + \theta)/B (\alpha, \beta)$.
Also in (\ref{1.2}),
\begin{eqnarray}
B_n \left( {\bf r} \right) =
\prod_{i = 1}^k B \left( r_i - r_{i - 1}, n - r_i + 1 \right).
\label{2.4}
\end{eqnarray}
\end{lemma}

\begin{note}
Since $B (\alpha, \beta) = \infty$ for $Re \beta \leq 0$, for (\ref{2.2})
to be finite we need $n - r_i + 1 + Re \bar{\boldsymbol{\theta}} > 0$ for $1 \leq i \leq k$.
\end{note}

\noindent
{\bf Proof of Lemma 2.1}
Set $I_k = LHS (\ref{2.2}) = \int B_n ({\bf u} : {\bf r}) \prod_{i = 1}^k (1 - u_i)^{\theta_i} du_1 \cdots du_k$
integrated over $0 < u_1 < \cdots < u_k < 1$ by (\ref{1.1}).
So, (\ref{2.2}), (\ref{2.4}) hold for $k = 1$.
Set $s_i = (u_i - u_{i - 1})/(1 - u_{i - 1})$.
Then
\begin{eqnarray*}
I_2 = \int_0^1 u_1^{r_1 - 1} \left( 1 - u_1 \right)^{\theta_1}
\int_{u_1}^1 \left( u_2 - u_1 \right)^{r_2 - r_1 - 1} \left( 1 - u_2 \right)^{r_3 - r_2 - 1 + \theta_2} d u_2/B_n ({\bf r}),
\end{eqnarray*}
which is the RHS (\ref{2.2}) with denominator replaced by the RHS (\ref{2.3}).
Putting $\boldsymbol{\theta} = {\bf 0}$ gives (\ref{2.2}), (\ref{2.4}) for $k = 2$.
Now use induction.
$\ \Box$

\begin{lemma}
In Lemma 2.1, the restriction
\begin{eqnarray}
1 \leq r_1 < \cdots < r_k \leq n
\mbox{ may be relaxed to }
1 \leq r_1 \leq \cdots \leq r_k \leq n.
\label{2.5}
\end{eqnarray}
\end{lemma}

\noindent
{\bf Proof}
For $k = 2$, the second factor in RHS (\ref{2.3})
is $b (r_2 - r_1, n - r_2 + 1 : \bar{\theta}_2) = f (\bar{\theta}_2)/f (0)$,
where $f (\bar{\theta}_2) = \Gamma (n - r_2 + 1 + \bar{\theta}_2)/\Gamma (n - r_1 + 1 + \bar{\theta}_2) = 1$ if $r_2 = r_1$
and the first factor is $b (r_1, n - r_1 + 1 : \bar{\theta}_1) = E (1 - U_{n r_1})^{\bar{\theta}_1}$.
Similarly, if $r_i = r_{i - 1}$, the $i$th factor is 1 and the product of the others is
$E \prod_{j = 1, j \neq i}^k (1 - U_{n r_j})^{\theta_j^{*}}$,
where $\theta_j^{*} = \theta_j$ for $j \neq i - 1$ and $\theta_j^{*} = \theta_{i - 1} + \theta_i$ for $j = i - 1$.
$\ \Box$

\begin{corollary}
In any formulas for $E g (X_{n, {\bf r}})$ for some function $g$, (\ref{2.5}) holds.
In particular it holds for the moments and cumulants of $X_{n, {\bf r}}$.
\end{corollary}

This result is very important as it means we can dispense with treating the $2^{k - 1}$ cases
$(r_i < r_{i + 1}$ or $r_i = r_{i + 1}$, $1 \leq i \leq k - 1$ separately.
For example, Hall (1978) treats the two cases for $\cos (X_{n, {\bf r}}, X_{n, {\bf s}})$ separately
and David and Johnson (1954) treat the $2^{k - 1}$ cases for the $k$th order cumulants
of $X_{n, {\bf r}}$ separately for $k \leq 4$.

\begin{theorem}
Under the conditions of Theorem 2.1,
\begin{eqnarray}
E \prod_{i = 1}^k X_{n, r_i}^{\theta_i} =
\sum_{i_1, \ldots, i_k = 0}^\infty
C_{i_1, \psi_1} \cdots C_{i_k, \psi_k} b_n \left( {\bf r} : \bar{\bf i} a - \bar{\boldsymbol{\theta}}/\alpha \right)
\label{2.6}
\end{eqnarray}
with $b_n$ as in (\ref{2.3}).
All terms are finite if $Re \bar{\boldsymbol{\theta}} < ({\bf s} + 1) \alpha$, where $s_i = u - r_i$.
\end{theorem}

\begin{lemma}
For $\alpha$, $\beta$ positive integers $\theta$ in $C$,
\begin{eqnarray}
b (\alpha, \beta : \theta) = \prod_{j = \beta}^{\alpha + \beta - 1}
\left( 1 + \theta/j \right)^{-1}.
\label{2.7}
\end{eqnarray}
So, for $\boldsymbol{\theta}$ in $C^k$,
\begin{eqnarray}
b_n ({\bf r} : \bar{\boldsymbol{\theta}}) = \prod_{i = 1}^k \prod_{j = s_i + 1}^{s_{i - 1}}
\left( 1 + \bar{\boldsymbol{\theta}}/j \right)^{-1},
\label{2.8}
\end{eqnarray}
where $s_i = n - r_i$ and $r_0 = 0$.
\end{lemma}

\noindent
{\bf Proof:}
LHS (\ref{2.7}) $= \Gamma (\beta + \theta) \Gamma (\alpha + \beta)/\{ \Gamma (\beta + \theta + \alpha) \Gamma (\beta) \}$.
But $\Gamma (\alpha + x)/\Gamma (x) = (x)_\alpha$, so (\ref{2.7}) holds, and hence (\ref{2.8}).
$\ \Box$

From (\ref{2.3}) we have, interpreting $\prod_{i = 2}^k b_i$ as 1 when $k - 1$,

\begin{lemma}
For $s_i = n - r_i$,
\begin{eqnarray}
b_n ({\bf r} : \bar{\boldsymbol{\theta}}) = B ({\bf s} : \bar{\boldsymbol{\theta}}) n!/\Gamma \left( n + 1 + \bar{\theta}_1 \right),
\label{2.9}
\end{eqnarray}
where
\begin{eqnarray*}
B ({\bf s} : \bar{\boldsymbol{\theta}}) = \Gamma \left( s_1 + 1 + \bar{\theta}_1 \right) \left( s_1 ! \right)^{-1}
\prod_{i = 2}^k b \left( s_{i - 1} - s_i, s_i + 1 : \bar{\theta}_1 \right)
\end{eqnarray*}
does not depend on $n$ for fixed ${\bf s}$.
\end{lemma}

\begin{lemma}
We have
\begin{eqnarray*}
n!/\Gamma (n + 1 + \theta) = n^{-\theta} \sum_{i = 0}^\infty e_i (\theta) n^{-i},
\end{eqnarray*}
where
\begin{eqnarray*}
&&
e_0 (\theta) = 1,\
e_1 (\theta) = -(\theta)_2/2,\
e_2 (\theta) = (\theta)_3 (3 \theta + 1)/24,
\\
&&
e_3 (\theta) = -(\theta)_4 (\theta)_2/(4! 2),\
e_4 (\theta) = (\theta)_5 (15 \theta^3 + 30 \theta^2 + 5 \theta - 2)/(5! 48),
\\
&&
e_5 (\theta) = -(\theta)_6 (\theta)_2 (3 \theta^2 + 7 \theta - 2)/(6! 16),
\\
&&
e_6 (\theta) = (\theta)_7 (63 \theta^5 + 315 \theta^4 + 315 \theta^3 - 91 \theta^2 - 42 \theta + 16)/(7! 576),
\\
&&
e_7 (\theta) = -(\theta)_8 (\theta)_2 (9 \theta^4 + 54 \theta^3 + 51 \theta^2 - 58 \theta + 16)/(8! 144).
\end{eqnarray*}
\end{lemma}

\noindent
{\bf Proof:}
Apply equation (6.1.47) of Abramowitz and Stegun (1964) for $i \leq 2$ and Withers and Nadarajah (2007b) for $i \leq 7$.
$\ \Box$

So, (\ref{2.6}), (\ref{2.9}) yield the joint moments of $X_{n, {\bf r}} n^{-1/\alpha}$
for fixed ${\bf s}$ as a power series in $(1/n, n^{-\alpha})$:

\begin{corollary}
We have
\begin{eqnarray}
E \prod_{i = 1}^k X_{n, n - s_i}^{\theta_i} =
\sum_{j = 0}^\infty n! \Gamma \left( n + 1 + j a - \bar{\psi}_1 \right)^{-1}
C_j \left( {\bf s} : \boldsymbol{\psi} \right),
\label{2.11}
\end{eqnarray}
where $\boldsymbol\psi = \boldsymbol\theta/\alpha$ and
\begin{eqnarray*}
C_j \left( {\bf s} : \boldsymbol{\psi} \right) = \sum \left\{
C_{i_1, \psi_1} \cdots  C_{i_k, \psi_k} B \left( {\bf s} : \bar{\bf i} a - \bar{\boldsymbol\psi} \right) :
i_1 + \cdots + i_k = j \right\}.
\end{eqnarray*}
\end{corollary}

So, if ${\bf s}$, $\boldsymbol\theta$ are fixed as $n \rightarrow \infty$ and $Re (\bar{\boldsymbol{\theta}}) < ({\bf s} + {\bf 1}) \alpha$,
\begin{eqnarray}
LHS (\ref{2.11}) = n^{\psi_1} \sum_{i, j = 0}^\infty
n^{-i - j a} e_i \left( j a - \bar{\psi}_1 \right) C_j
\left( {\bf s} : \boldsymbol{\psi} \right).
\label{2.12}
\end{eqnarray}
If $a$ is rational, say $a = M/N$ then
\begin{eqnarray}
LHS (\ref{2.11}) = n^{\bar{\psi}_1} \sum_{m = 0}^\infty
n^{-m/N} d_m \left( {\bf s} : \boldsymbol\psi \right),
\label{2.13}
\end{eqnarray}
where
\begin{eqnarray}
d_m \left( {\bf s} : \boldsymbol\psi \right)
&=&
\sum \left\{ e_i \left(  j a - \bar{\psi}_1 \right)
C_j \left( {\bf s} : \boldsymbol{\psi} \right) : i N + j M = m \right\}
\nonumber
\\
&=&
\sum \left\{ e_{m - j a} \left(  j a - \bar{\psi}_1 \right)
C_j \left( {\bf s} : \boldsymbol{\psi} \right) : 0 \lq j \leq m/a \right\}
\nonumber
\end{eqnarray}
if $N = 1$; so for $d_m$ to depend on $c_1$ and not just $c_0$ we need $m \leq M$.

\begin{note}
The following dimensional checks can be used throughout.
By (\ref{1.5}), $dim c_i = (dim X)^{\alpha + i \beta}$.
By (\ref{2.1}), $dim C_{i \psi} = (dim X)^{\theta}$.
Also $dim \bar{x}_i = (dim X)^{-\alpha}$ and
$dim d_m (s : \psi) = dim C_j (s : \psi) = (dim X)^{\bar{\theta}_1}$.
\end{note}

\begin{note}
The leading term in (\ref{2.12}) does not involve $c_1$ so may be deduced from the
multivariate extreme value distribution that the law of $X_{n, n - s_i}$, suitably normalized, tends to.
The same is true of the leading terms of its cumulants.
See Withers and Nadarajah (2007a) for details.
\end{note}

The leading terms in (\ref{2.12}) are
\begin{eqnarray*}
n^{\bar{\psi}_1} \left[ \left\{ 1 - n^{-1} <\bar{\psi}_1>_2/2 \right\} C_0 ({\bf s} : \boldsymbol\psi) +
n^{-a} C_0 ({\bf s} : \boldsymbol\psi) + O \left( n^{-2 a_0} \right) \right],
\end{eqnarray*}
where
\begin{eqnarray*}
a_0
&=&
\min (a, 1),
\\
C_0 ({\bf s} : \boldsymbol\psi)
&=&
c_0 B ({\bf s} : -\bar{\boldsymbol\psi}),
\\
C_1 ({\bf s} : \boldsymbol\psi)
&=&
c_0^{\bar{\psi}_1 - a - 2} c_1 \sum_{j = 1}^k \psi_j B \left( {\bf s} : a {\bf I}_j - \bar{\boldsymbol\psi} \right)
\end{eqnarray*}
and for ${\bf I}_j = \bar{\bf i}$ for $i_m = \delta_{m j}$, that is $I_{jm} = I (m \leq j)$.
For $k = 1$,
\begin{eqnarray*}
C_j (s : \psi)
&=&
C_{j \psi} (s + 1)_{j a - \psi},
\\
C_0 (s : \psi)
&=&
c_0^\psi (s + 1)_{-\psi} = c_0^\psi/<s>_\psi,
\\
C_1 (s : \psi)
&=&
\psi c_0^{\psi - a - 1} c_1 (s + 1)_{a - \psi} = \psi c_0^{\psi - a - 1} c_1/<s>_{\psi - a}.
\end{eqnarray*}
Set $\pi_{\bf s} (\lambda) = b (s_1 - s_2, s_2 + 1 : \lambda) = \prod_{j = s_2 + 1}^{s_1} 1/(1 + \lambda/j)$ for $\lambda$ an integer.
For example, $\pi_{\bf s} (1) = (s_2 + 1)/(s_1 + 1)$ and $\pi_{\bf s} (-1) = s_1/s_2$.
Then for $k = 2$,
\begin{eqnarray*}
C_0 ({\bf s} : \lambda {\bf 1})
&=&
c_0^{2 \lambda} <s_1>_{2 \lambda}^{-1} \pi_{\bf s} (-\lambda)
\\
&=&
c_0^2 \left( s_1 - 1 \right)^{-1} s_2 \mbox{ for $\lambda = 1$}
\\
&=&
c_0^2 <s_2 - 2>_2^{-1} <s_2>_2^{-1} \mbox{ for $\lambda = 2$}
\end{eqnarray*}
and
\begin{eqnarray*}
C_1 ({\bf s} : \lambda {\bf 1})
&=&
\lambda c_0^{2 \lambda - a - 1} c_1 <s_1>_{2 \lambda - a}^{-1} \left\{ \pi_{\bf s} (-\lambda) + \pi_{\bf s} (a - \lambda) \right\}
\\
&=&
\lambda c_0^{1 - a} c_1 <s_1>_{2 - a}^{-1} \left\{ s_1/s_2 + \pi_{\bf s} (a - 1) \right\} \mbox{ for $\lambda = 1$}
\\
&=&
\lambda c_0^{3 - a} c_1 <s_1>_{4 - a}^{-1} \left\{ <s_1>_2 <s_2>_2^{-1} + \pi_{\bf s} (a - 2) \right\} \mbox{ for $\lambda = 2$.}
\end{eqnarray*}
Set $\lambda = 1/\alpha$, $Y_{ns} = X_{n, n - s}/(n c_0)^\lambda$ and $E_c = \lambda c_0^{-a - 1} c_1$.
Then for $s > \lambda - 1$
\begin{eqnarray}
E Y_{ns} = \left\{ 1 - n^{-1} <\lambda>_2/2 \right\} <s>_\lambda^{-1} +
n^{-a} E_c <s>_{\lambda - a}^{-1} + O \left( n^{-2 a_0} \right)
\label{2.15}
\end{eqnarray}
and for $s_1 > 2 \lambda - 1$, $s_2 > \lambda - 1$, $s_1 \geq s_2$,
\begin{eqnarray}
E Y_{ns_1} Y_{ns_2} = \left\{ 1 - n^{-1} <2 \lambda>_2/2 \right\} B_{20} + n^{-a} E_c D_a + O \left( n^{-2 a_0} \right),
\label{2.17}
\end{eqnarray}
where $B_{20} = <s_1>^{-1}_{2 \lambda} \pi_{\bf s} (-\lambda)$, $D_a = <s_1>_{2 \lambda - a}^{-1} \{ \pi_{\bf s} (-\lambda) + \pi_{\bf s} (a - \lambda) \}$ and
\begin{eqnarray}
Covar \left( Y_{ns_1}, Y_{ns_2} \right) = F_0 + F_1/n + E_c F_2/n + O \left( n^{-2 a_0} \right),
\label{2.18}
\end{eqnarray}
where $F_0 = B_{20} - <s_1>_\lambda^{-1} <s_2>_\lambda^{-1}$,
$F_1 = <\lambda>_2 <s_1>_\lambda^{-1} <s_2>_\lambda^{-1} - <2 \lambda>_2 B_{20}/2$
and $F_2 = D_a - <s_1>_\lambda^{-1} <s_2>_{\lambda - a}^{-1} - <s_1>_{\lambda - a}^{-1} <s_2>_{\lambda}^{-1}$.
Similarly, we may use (\ref{2.12}) to approximate higher order cumulants.
If $a = 1$ this gives $E Y_{ns}$ and $Covar (Y_{ns_1}, Y_{ns_2})$ to $O(n^{-2})$.

\begin{example}
Suppose $\alpha = 1$.
Then $Y_{ns} = X_{n, n - s}/(n c_0)$, $E_c = c_0^{-a - 1} c_0$, $B_{20} = -F_1 = (s_1 - 1)^{-1} s_2^{-1}$,
$F_0 = <s_1>_2^{-1} s_2^{-1}$, $D_a = <s_1>_{2 - a}^{-1} G_a$,
where $G_a = s_1 s_2^{-1} + \pi_{\bf s} (a - 1)$ for $s_1 \geq s_2$,
$G_a = 2$ for $s_1 = s_2$ and $F_2 = D_a - s_1^{-1} <s_2>_{1 - a}^{-1} - s_2^{-1} <s_1>_{1 - a}^{-1}$.
So,
\begin{eqnarray}
E Y_{ns} = s^{-1} + n^{-a} E_c <s>_{1 - a}^{-1} + O(n^{-2 a_0})
\label{2.19}
\end{eqnarray}
for $s > 0$ and (\ref{2.17})-(\ref{2.18}) hold if
\begin{eqnarray}
s_1 > 1,\
s_2 > 0,\
s_1 \geq s_2.
\label{2.20}
\end{eqnarray}
A little calculation shows that $C_0 ({\bf s} : {\bf 1}) = c_0^k B_{k0}$,
$C_1 ({\bf s} : {\bf 1}) = c_0^{k - a - 1} c_1 B_{k \cdot}$, and
\begin{eqnarray*}
E \prod_{i = 1}^k Y_{n, s_i}
&=&
\left\{ 1 + n^{-1} <k>_2/2 \right\} B_{k0} + n^{-a} E_c B_k + O(n^{-2 a_0})
\nonumber
\\
&=&
m_0 (s) + n^{-1} m_1 (s) + n^{-a} m_a (s) + O(n^{-2 a_0})
\end{eqnarray*}
say for $s_i > k - i$, $1 \leq i \leq k$ and $s_1 \geq \cdots \geq s_k$, where
\begin{eqnarray*}
B_{k \cdot}
&=&
\sum_{j = 1}^k B_{kj},
\\
B_{k0}
&=&
\prod_{i = 1}^k 1/\left( s_1 - k + 1 \right),
\\
B_{kj}
&=&
\prod_{i = 1}^{j - 1} \left( s_i - k + a + i \right)^{-1}
<s_j - k + j + 1>_{a - 1}
\prod_{i = j + 1}^k \left( s_i - k + i \right)^{-1},
\\
B_{kk}
&=&
\prod_{i = 1}^{k - 1} \left( s_i - k + a + i \right)^{-1} <s_k>_{1 - a}^{-1}
\end{eqnarray*}
for $s_i > k - i$ and $1 \leq j < k$.
For example, $B_{10} = s_1$, $B_{20} = (s_1 - 1)^{-1} s_2^{-1}$ and $B_{30} = (s_1 - 2)^{-1} (s_2 - 1)^{-1} s_3^{-1}$.
So, $\kappa_n (s) = \kappa (Y_{n s_1}, \ldots, Y_{n s_k})$ is given by
$\kappa_n (s = \kappa_0 (s) + n^{-1} \kappa_1 (s) + n^{-a} \kappa_a (s) + O(n^{-2 a_0})$, where, for example,
writing $\Sigma^3 a(s_1) b(s_2 s_3) = a(s_1) b(s_2 s_3) + a(s_2) b(s_3 s_1) + a(s_3) b(s_1 s_2)$,
\begin{eqnarray*}
\kappa_0 \left( s_1 s_2 s_3 \right)
&=&
m_0 \left( s_1 s_2 s_3 \right) - \sum^3 m_0 \left( s_1 \right) m_0 \left( s_2 s_3 \right) +
2 \prod_{i = 1}^3 m_0 \left( s_i \right)
\\
&=&
2 \left( s_1 + s_2 - 2 \right) D \left( s_1 s_2 s_3 \right),
\\
\kappa_1 \left( s_1 s_2 s_3 \right)
&=&
m_1 \left( s_1 s_2 s_3 \right) - \sum^3 m_0 \left( s_1 \right) m_1 \left( s_2 s_3 \right)
\\
&=&
2 \left\{ s_2 \left( 1 - 2 s_1 \right) + s_1 - s_1^2 \right\}/D \left( s_1 s_2 s_3 \right)
\mbox{ since $m_1 (s_1) = 0$,}
\\
\kappa_a \left( s_1 s_2 s_3 \right)
&=&
m_a \left( s_1 s_2 s_3 \right) - \sum^3 \left\{ m_0 \left( s_1 \right) m_a \left( s_2 s_3 \right) +
m_a \left( s_1 \right) m_0 \left( s_2 s_3 \right) \right\}
\\
&&
\quad +
2 \sum^3 m_0 \left( s_1 \right) m_0 \left( s_2 \right) m_a \left( s_3 \right),
\end{eqnarray*}
where $D (s_1 s_2 s_3) = <s_1>_3 <s_2>_2 s_3$.

Consider the case $a = 1$.
Then $\kappa_a (s_1 s_2 s_3) = 0$ so
\begin{eqnarray}
\kappa_n \left( s_1 s_2 s_3 \right)
&=&
2 \left\{ s_1 + s_2 - 2 + n^{-1} \left( s_2 \left( 1 - 2 s_1 \right) +
s_1 - s_1^2 \right) \right\}/D \left( s_1 s_2 s_3 \right) + O \left( n^{-2} \right).
\label{2.23}
\end{eqnarray}
Set $s_{\cdot} = \sum_{j = 1}^k s_j$.
Then
\begin{eqnarray*}
&&
B_{1 \cdot} = B_{11} - 1,\
B_{22} = 1/s_2,\
B_{22} = 1/s_2,\
B_{22} = s_1,
\\
&&
B_{2 \cdot} = s_1^{-1} + s_2^{-1} = \left( s_1 + s_2 \right)/\left( s_1 s_2 \right),
\\
&&
B_{31} = \left( s_2 - 1 \right)^{-1} s_3^{-1},\
B_{32} = \left( s_1 - 1 \right)^{-1} s_3^{-1},\
B_{33} = \left( s_1 - 1 \right)^{-1} s_2^{-1},
\\
&&
B_{3 \cdot} = \left\{ s_2 \left( s_\cdot - 2 \right) - s_3 \right\}
\left( s_1 - 1 \right)^{-1} <s_2>_2^{-1} s_3^{-1},
\\
&&
B_{41} = \left( s_2 - 2 \right)^{-1} \left( s_3 - 1 \right)^{-1} s_4^{-1},\
B_{42} = \left( s_1 - 2 \right)^{-1} \left( s_3 - 1 \right)^{-1} s_4^{-1},
\\
&&
B_{43} = \left( s_1 - 2 \right)^{-1} \left( s_2 - 1 \right)^{-1} s_4^{-1},\
B_{44} = \left( s_1 - 2 \right)^{-1} \left( s_2 - 1 \right)^{-1} s_3^{-1},
\\
&&
B_{4 \cdot} = \left\{ s_\cdot s_3 \left( s_2 - 2 \right) + s_3 \left( s_2 - 4 s_2 + 4 \right) - s_2 s_4 \right\}
\left\{ \left( s_1 - 2 \right) <s_2 - 2>_2 <s_3>_2 s_4 \right\}^{-1}.
\end{eqnarray*}
Also $E_c = c_0^{-2} c_1$, $D_a = s_1^{-1} + s_2^{-1}$, $F_2 = 0$, and
\begin{eqnarray}
E Y_{ns}
&=&
s^{-1} + n^{-1} E_c + O \left( n^{-2} \right)
\mbox{ for $s > 0$,}
\label{2.24}
\\
E Y_{n, s_1} Y_{n, s_2}
&=&
\left( 1 - n^{-1} \right) B_2 + n^{-1} E_c D_a + O \left( n^{-2} \right)
\mbox{ if (\ref{2.20}) holds,}
\label{2.25}
\\
Covar \left( Y_{n, s_1}, Y_{n, s_2} \right)
&=&
<s_1>_2^{-1} s_2^{-1} \left( s - n^{-1} s_1 \right) + O \left( n^{-2} \right)
\mbox{ if (\ref{2.20}) holds.}
\label{2.26}
\end{eqnarray}

In the case $a \geq 2$, (\ref{2.24})-(\ref{2.26}) hold with $E_c$ replaced by $0$.
In the case $a \leq 1$, (\ref{2.17})-(\ref{2.19}) with $a_0 = a$ give terms $O (n^{-2 a})$
with the $n^{-1}$ terms disposable if $a \leq 1/2$.
\end{example}

We now investigate what extra terms are needed to make
(\ref{2.24})-(\ref{2.26}) depend on $c$ when $a = 1$ or $2$.

\begin{example}
$\alpha = \beta = 1$.
Here, we fine the coefficients of $n^{-2}$.
By (\ref{2.13}),
\begin{eqnarray*}
d_2 ({\bf s} : \boldsymbol{\psi})
&=&
\sum_{j = 0}^2 e_{2 - j}
\left( j - \bar{\psi}_1 \right) C_j ({\bf s} : \boldsymbol{\psi}) +
e_2 \left( -\bar{\psi}_1 \right) C_0 ({\bf s} : \boldsymbol{\psi})
\\
&&
\quad +
\bar{e}_1 \left( 1 - \bar{\psi}_1 \right) C_1 ({\bf s} : \boldsymbol{\psi}) + C_2 ({\bf s} : \boldsymbol{\psi})
\\
&=&
C_2 ({\bf s} : \boldsymbol{\psi})
\mbox{ if $\bar{\psi}_1 = 1$ or $2$.}
\end{eqnarray*}
For $k = 1$, $C_2 (s : \psi) = C_{2 \psi} (s + 1)_{2 - \psi}$,
where $C_{2 \psi} = \psi c_0^{\psi - 4} \{ c_0 c_2 + (\psi - 3) c_1^2/2 \}$,
so $d_2 (s : 1) = (s + 1) F_c$, where $F_c = c_0^{-3} (c_0 c_2 - c_1^2)$,
so in (\ref{2.24}) we may replace $O (n^{-2})$ by $n^{-2} (s + 1) F_c c_0^{-1} + O (n^{-3})$.
For $k = 2$,
\begin{eqnarray*}
C_2 ({\bf s} : {\bf 1})
&=&
\sum \left\{ C_{i1} C_{j1} B ({\bf s} : 0, j - 1) : i + j = 2 \right\}
\\
&=&
C_{01} C_{21} \left\{ B ({\bf s} : 0, 1) + B ({\bf s} : 0, -1) \right\} + C_{11}^2 B ({\bf s} : {\bf 0}),
\end{eqnarray*}
where $B ({\bf s} : 0, \lambda) = b (s_1 - s_2, s_2 + 1 : \lambda) = \pi_{{\bf s}} (\lambda)$, so
$d_2 ({\bf s} : {\bf 1}) = C_2 ({\bf s} : {\bf 1}) - D_{2, {\bf s}} H_c + c_0^{-2} c_1^2$,
where $D_{2, {\bf s}} = (s_2 + 1) (s_1 + 1)^{-1} + s_1 s_2^{-1}$,
$H_c = c_0^{-2} (c_0 c_2 - c_1^2)$ and in (\ref{2.25}) we may replace $O (n^{-2})$
by $n^{-2} d_2 ({\bf s} : {\bf 1}) c_0^{-2} + O(n^{-3})$.
Upon simplifying this gives
\begin{eqnarray*}
Covar \left( Y_{n, s_1}, Y_{n, s_2} \right) =
<s_1>_2^{-1} s_2^{-1} \left( 1 - n^{-1} s_1 \right) - c_0^{-2} H_c F_{3, {\bf s}} n^{-2} + O \left( n^{-2} \right),
\end{eqnarray*}
where $F_{3, {\bf s}} = (s_2 + 1)/<s_1>_2 + s_2^{-1}$.
\end{example}

\begin{example}
$\alpha = 1$, $\beta = 2$.
So, $a = 2$, $\lambda = 1$, $\boldsymbol\psi = \boldsymbol\theta$.
By (\ref{2.13}),
\begin{eqnarray*}
d_2 ({\bf s} : \boldsymbol{\psi})
&=&
\sum_{j = 0}^1 e_{2 - 2j} \left( 2 j - \bar{\psi}_1 \right) C_j ({\bf s} : \boldsymbol{\psi})
\\
&=&
e_2 \left( -\bar{\psi}_1 \right) C_0 ({\bf s} : \boldsymbol{\psi}) + C_1 ({\bf s} : \boldsymbol{\psi})
\\
&=&
C_1 ({\bf s} : \boldsymbol{\psi})
\mbox{ if $\bar{\psi}_1 = 0, 1$ or $2$.}
\end{eqnarray*}
For $k = 1$,
\begin{eqnarray*}
C_1 (s : \psi) = \psi c_0^{\psi - 3} c_1 <s>_{\psi - 2}^{-1} =
\left\{
\begin{array}{ll}
c_0^{-2} c_1 (s + 1), & \mbox{if $\psi = 1$,}\\
2 c_0^{-1} c_1, & \mbox{if $\psi = 2$,}
\end{array}
\right.
\end{eqnarray*}
so $E Y_{ns} = s^{-1} + c_0^{-3} c_1 (s + 1) n^{-2} + O (n^{-3})$ for $s > 0$.
For $k = 2$, $C_1 ({\bf s} : {\bf 1}) = c_0^{-1} c_1 D_{2, {\bf s}}$ for $D_{2, {\bf s}}$ above, so
\begin{eqnarray*}
E Y_{n, s_1} Y_{n, s_2} = \left( 1 - n^{-1} \right) \left( s_1 - 1 \right)^{-1} s_2^{-1} + n^{-2} c_0^{-3} c_1 D_{2, {\bf s}} + O \left( n^{-3} \right)
\end{eqnarray*}
and
\begin{eqnarray*}
Covar \left( Y_{n, s_1}, Y_{n, s_2} \right) =
<s_1>_2^{-1} s_2^{-1} \left( 1 - n^{-1} s_1 \right) - n^{-2} c_0^{-3} c_1 F_{3, {\bf s}} + O \left( n^{-3} \right).
\end{eqnarray*}
\end{example}

\section{Examples}

\begin{example}
For Student's $t$ distribution, $X = t_N$ has density
\begin{eqnarray*}
\left( 1 + x^2/N \right)^{-\gamma} g_N = \sum_{i = 0}^\infty d_i x^{-2 \gamma - 2 i},
\end{eqnarray*}
where $\gamma = (N + 1)/2$, $g_N = \Gamma (\gamma)/\{ \sqrt{N \pi} \Gamma (N/2) \}$ and $d_i = {-\gamma \choose i} N^{\gamma + i} g_N$.
So, (\ref{1.5}) holds with $\alpha = N$, $\beta = 2$ and $c_i = d_i/(N + 2i)$:
\begin{eqnarray*}
c_0
&=&
N^{\gamma - 1} g_N,
\\
c_1
&=&
-\gamma N^{\gamma + 1} (N + 2)^{-1} g_N = -N^{\gamma + 1} (N + 1) (N + 2)^{-1} g_N/2,
\\
c_2
&=&
(\gamma)_2 N^{\gamma + 2} (N + 4)^{-1} g_N/2,
\\
c_3
&=&
-(\gamma)_3 N^{\gamma + 3} G_N (N + 6)^{-1}/6,
\end{eqnarray*}
and so on.
So, $a = 2/N$ and (\ref{2.13}) gives an expression in powers of $n^{-a/2}$ if $N$ is odd or $n^{-a}$ if $N$ is even.
The first term in (\ref{2.13}) to involve $c_1$, not just $c_0$, is the coefficient of $n^{-a}$.
\end{example}

Putting $N = 1$ we get

\begin{example}
For the Cauchy distribution, (\ref{1.5}) holds with $\alpha = 1$, $\beta = 2$ and $c_i = (-1)^i (2 i + 1)^{-1} \pi^{-1}$.
So, $a = 2$, $\psi = \theta$, $C_{0 \psi} = \pi^{-\psi}$, $C_{1 \psi} = -\psi \pi^{2 - \psi}/3$,
$C_{2 \psi} = \psi \pi^{4 - \psi} \{ 1/5 + (\psi - 5)/a \}$ and
$C_{3 \psi} = -\psi \pi^{6 - \psi} \{ 1/105 - 2 \psi/15 + (\psi + 1)_2/162 \}$.
By Example 2.3, $Y_{ns} = (\pi/n) X_{n, n - s}$ satisfies
\begin{eqnarray}
E Y_{ns} = s^{-1} - n^{-2} \pi^2 (s + 1) + O \left( n^{-3} \right)
\label{3.1}
\end{eqnarray}
for $s > 0$ and when (\ref{2.20}) holds
\begin{eqnarray}
E Y_{n, s_1} Y_{n, s_2} = \left( 1 - n^{-1} \right) \left( s_1 - 1 \right)^{-1} s_2^{-1} - n^{-2} \pi^2 D_{2, {\bf s}}/3 + O \left( n^{-3} \right)
\label{3.2}
\end{eqnarray}
for $D_{2, {\bf s}} = (s_2 + 1)/(s_1 + 1) + s_1/s_2$ and
\begin{eqnarray*}
Covar \left( Y_{n, s_1}, Y_{n, s_2} \right) =
<s_1>_2^{-1} s_2^{-1} \left( 1 - n^{-1} s_1 \right) + n^{-2} \pi^2 F_{3, {\bf s}}/3 + O \left( n^{-3} \right)
\end{eqnarray*}
for $F_{3, {\bf s}} = (s_2 + 1)/<s_1>_2 + s_2^{-1}$.
Hall (1978, page 274) gave the first term in (\ref{3.1}) and (\ref{3.2}) when $s_1 = s_2$
but his version of (\ref{3.2}) for $s_1 > s_2$ replaces $(s_1 - 1)^{-1} s_2^{-1}$
and $D_{2, {\bf s}}$ by complicated expressions each with $s_1 - s_2$ terms.
The joint order of order three for $\{ Y_{n, s_i}, 1 \leq i \leq 3 \}$ is given by (\ref{2.23}).
Hall points out that $F^{-1} (u) = \cot (\pi - \pi u)$, so
$F^{-1} (u) = \sum_{i = 0}^\infty (1 - u)^{2 i - 1} C_{i1}$,
where $C_{i1} = (-4 \pi^2)^i \pi^{-1} B_{2i}/(2i)!$.
Note 2.1 could be used.
We have not done so.
\end{example}

\begin{example}
Consider the $F$ distribution.
For $N, M \geq 1$, set $\nu = M/N$, $\gamma = (M + N)/2$ and $g_{MN} = \nu^{M/2}/B (M/2, N/2)$.
Then $X = F_{M, N}$ has density
\begin{eqnarray*}
x^{M/2} \left( 1 + \nu x \right)^{-\gamma} g_{MN} =
\nu^{-\gamma} x^{-N/2} \left( 1 + \nu^{-1} x^{-1} \right)^{-\gamma} g_{MN} = \sum_{i = 0}^\infty d_i x^{-N/2 - i},
\end{eqnarray*}
where $d_i = h_{MN} {-\gamma \choose i} \nu^i$ and $h_{MN} = g_{MN} \nu^{-\gamma} = \nu^{-N/2}/B (M/2, N/2)$.
So, for $N > 2$, (\ref{2.1}) holds with $\alpha = N/2 - 1$, $\beta = 1$ and $c_i = d_i/(N/2 + i - 1)$.
If $N = 4$ then $\alpha = 1$ and Examples 2.1-2.2 apply.
Otherwise (\ref{2.15})-(\ref{2.18}) give $E Y_{n, {\bf s}}$, $E Y_{n, s_1} Y_{n, s_2}$
and $Covar (Y_{n, s_1}, Y_{n, s_2})$ to $O (n^{-2 a_0})$,
where $Y_{n, s} = X_{n, n - s}/(n c_0)\lambda$, $\lambda = 1/\alpha$, $a = 2/(N - 2)$,
$a_0 = \min (a, 1) = a$ if $N \geq 4$ and $a_0 = \min (a, 1) = 1$ if $N < 4$.
\end{example}

\begin{example}
Consider the stable laws.
Feller (1966, page 549) proves that the general stable law of index $\alpha \in (0, 1)$ has density
\begin{eqnarray*}
\sum_{k = 1}^\infty |x|^{-1 - a k} a_k (\alpha, \gamma),
\end{eqnarray*}
where $a_k (\alpha, \gamma) = (1/\pi) \Gamma (k \alpha + 1) \{ (-1)^k/k! \} \sin \{ k \pi (\gamma - \alpha)/2 \}$ and $\mid \gamma \mid \leq \alpha$.
So, for $x > 0$ its distribution $F$ satisfies (\ref{2.1}) with $\beta = \alpha$
and $c_i = a_{i + 1} (\alpha, \gamma) \gamma^{-1} (i + 1)^{-1}$.
Since $a = 1$ the first two moments of $Y_{n, s} = X_{n, n - s}/(n c_0)^\lambda$,
where $\lambda = 1/\alpha$ are given to $O(n^{-2})$ by (\ref{2.15})-(\ref{2.18}).
\end{example}

\begin{example}
Finally, consider the second extreme value distribution.
Suppose $F (x)$ $=$ $\exp$ $(-x^{-\alpha})$ for $x > 0$, where $\alpha > 0$.
Then (\ref{1.5}) holds with $\beta = \alpha$ and $c_i = (-1)^i/(i + 1)!$.
Since $a = 1$ the first two moments of $Y_{n, s} = X_{n, n - s}/n^{1/\alpha}$
are given to $O(n^{-2}$ by (\ref{2.15})-(\ref{2.18}).
\end{example}

\newpage

\section*{Appendix A: An Inversion Theorem}

Given $x_j = y_j/j!$ for $j \geq 1$ set
\begin{eqnarray}
S = \widehat{S} (t, x) = \sum_{j = 1}^\infty x_j t^j = S (t, y) = \sum_{j = 1}^\infty y_j t^j/j!.
\label{A1}
\end{eqnarray}
The partial ordinary and exponential Bell polynomials $\widehat{B}_{ri} (x)$ and $B_{ri} (y)$ are defined for $r = 0, 1, \ldots$ by
\begin{eqnarray*}
S^i = \sum_{r = i}^\infty t^r \widehat{B}_{ri} (x) = i! \sum_{r = i}^\infty t^r B_{ri} (y)/r!.
\end{eqnarray*}
So, $\widehat{B}_{r0} (x) = B_{r0} (y) = \delta_{r0}$ (1 or 0 as $r = 0$ or $r \neq 0$),
$\widehat{B}_{ri} (\lambda x) = \lambda^i \widehat{B}_{ri} (x)$ and $B_{ri} (\lambda y) = \lambda^i B_{ri} (y)$.
They are tabled on pages 307--309 of Comtet  (1974) for $r \leq 10$ and 12.
Note that
\begin{eqnarray}
(1 + \lambda S)^\alpha = \sum_{r = 0}^\infty t^r \widehat{C}_r = \sum_{r = 0}^\infty t^r C_r/r!,
\label{A3}
\end{eqnarray}
where
\begin{eqnarray}
\widehat{C}_r = \widehat{C}_r (\alpha, \lambda, x) =
\sum_{i = 0}^r \widehat{B}_{ri} (x) {\alpha \choose i} \lambda^i
\label{A4}
\end{eqnarray}
and
\begin{eqnarray*}
C_r = C_r (\alpha, \lambda, x) =
\sum_{i = 0}^r B_{ri} (y) <\alpha>_i \lambda^i.
\end{eqnarray*}
So, $\widehat{C}_0 = 1$, $\widehat{C}_1 = \alpha \lambda x_1$,
$\widehat{C}_2 = \alpha \lambda x_2 + <\alpha>_2 \lambda^2 x_1^2/2$,
$\widehat{C}_3 = \alpha \lambda x_3 + <\alpha>_2 \lambda^2 x_1 x_2 + <\alpha>_3 \lambda^3 x_1^3/6$
and $C_0 = 1$, $C_1 = \alpha \lambda y_1$, $C_2 = \alpha \lambda y_2 + <\alpha>_2 \lambda^2 y_1^2$.
Similarly,
\begin{eqnarray*}
\log (1 + \lambda S) = \sum_{r = 1}^\infty t^r \widehat{D}_r = \sum_{r = 1}^\infty t^r D_r/r!
\end{eqnarray*}
and
\begin{eqnarray*}
\exp (\lambda S) = 1 + \sum_{r = 1}^\infty t^r \widehat{B}_r = 1 + \sum_{r = 1}^\infty t^r B_r/r!,
\end{eqnarray*}
where
\begin{eqnarray*}
\widehat{D}_r = \widehat{D}_r (\lambda, x) = -\sum_{i = 1}^r \widehat{B}_{ri} (x) (-\lambda)^i/i!,
\end{eqnarray*}
\begin{eqnarray*}
D_r = D_r (\lambda, y) = -\sum_{i = 1}^r B_{ri} (y) (-\lambda)^i/(i - 1)!,
\end{eqnarray*}
\begin{eqnarray*}
\widehat{B}_r = \widehat{B}_r (\lambda, x) = \sum_{i = 1}^r \widehat{B}_{ri} (x) \lambda^i/i!
\end{eqnarray*}
and
\begin{eqnarray*}
B_r = B_r (\lambda, y) = \sum_{i = 1}^r B_{ri} (y) \lambda^i.
\end{eqnarray*}
Here, $\widehat{B}_r (1, x)$ and $B_r (1, y)$ are known as the {\it complete} ordinary and exponential Bell polynomials.
If $x_j = y_j = 0$ for $j$ even, then $S = t^{-1} \sum_{j = 1}^\infty X_j t^{2 j}$,
where $X_j = x_{2j - 1}$, so
\begin{eqnarray*}
S^i = t^{-i} \sum_{r = i}^\infty t^{2 r} \widehat{B}_{ri} (X)
\mbox{ and }
\exp (\lambda S) = 1 + \sum_{k = 1}^\infty t^k \widehat{B}_k,
\end{eqnarray*}
where
\begin{eqnarray*}
\widehat{B}_k = \sum \left\{ \widehat{B}_{ri} (X) \lambda^i/i! : i = 2 r - k, k/2 < r \leq k \right\}.
\end{eqnarray*}
The following derives from Lagrange's inversion formula.

\noindent
{\bf Theorem A.1}
{\it Let $k$ be a positive integer and $a$ any real number.
Suppose
\begin{eqnarray*}
v/u = \sum_{i = 0}^\infty x_i u^{ia} = \sum_{i = 0}^\infty y_i v^{ia}/i!
\end{eqnarray*}
with $x_0 \neq 0$.
Then
\begin{eqnarray*}
(u/v)^k = \sum_{i = 0}^\infty x_i^{*} v^{ia} = \sum_{i = 0}^\infty y_i^{*} v^{ia}/(ia)!,
\end{eqnarray*}
where $x_i^{*} = x_i^{*} (a, k, x)$ and $y_i^{*} = y_i^{*} (a, k, y)$ are given by
\begin{eqnarray}
x_i^{*} = k n^{-1} \widehat{C}_i \left( -n, 1/x_0, x \right) = k x_0^{-n}
\sum_{j = 0}^i (n + 1)_{j - 1} \widehat{B}_{ij} (x) \left( -x_0 \right)^{-j}/j!
\label{A15}
\end{eqnarray}
and
\begin{eqnarray}
y_i^{*} = k n^{-1} C_i \left( -n, 1/y_0, y \right) = k y_0^{-n}
\sum_{j = 0}^i (n + 1)_{j - 1} B_{ij} (y) \left( -y_0 \right)^{-j},
\label{A16}
\end{eqnarray}
respectively, where $n = k + ai$.}

\noindent
{\bf Proof:}
$u/v$ has a power series in $v^a$ so that $(u/v)^k$ does also.
A little work shows that (\ref{A15})-(\ref{A16}) are correct for $i = 0, 1, 2, 3$
and so by induction that $x_i^{*} x_0^{ia}$ and $y_i^{*} y_0^{ia}$ are
polynomials in $a$ of degree $i - 1$.
Hence, (\ref{A15})-(\ref{A16}) will hold true for all $a$ if they hold true for all positive integers $a$.
Suppose then $a$ is a positive integer.
Since $v/u = x_0 (1 + x_0^{-1} S)$ for $S = \widehat{S} (u^a, x) = S (u^a, y)$,
the coefficient of $u^{ai}$ in $(v/y)^{-n}$ is
$x_0^{-n} \widehat{C}_i (-n, 1/x_0, x) = y_0^{-n} C_i (-n, 1/y_0, y)/(n - k)!$.
Now set $n = k + ai$ and apply Theorem A in Comtet (1974, page 148) to
$v = f (u) = \sum_{i = 0}^\infty x_i u^{1 + a i}$.
$\ \Box$

\noindent
{\bf Note A.1}
{\it Comtet (1978, page 15, Theorem F) proves (\ref{A15}) for the case $k = 1$ and $a$ a positive integer.}


\begin{thebibliography}{999}


\bibitem{}
Abramowitz, M. and Stegun, I. A. (1964).
{\it Handbook of Mathematical Functions}.
National Bureau of Standards, Washington DC.



\bibitem{}
Comtet, L. (1974).
{\it Advanced Combinatorics}.
Reidel, Dordrecht.


\bibitem{}
David, F. N. and Johnson, N. L. (1954).
Statistical treatment of censored data.
Part I: Fundamental formulae.
{\it Biometrika}, {\bf 41}, 225--231.

\bibitem{}
Feller, W. (1966).
{\it An Introduction to Probability Thory and Its Applications}, volume 2.
John Wiley and Sons, New York.


\bibitem{}
Hall, P. (1978).
Some asymptotic expansions of moments of order statistics.
{\it Stochastic Processes and Their Applications}, {\bf 7}, 265--275.

\bibitem{}
James, G. S. and Mayne, A. J. (1962).
Cumulants of functions of random variables.
{\it Sankhy\=a}, A, {\bf 24}, 47--54.


\bibitem{}
McCord, J. R. (1964).
On asymptotic moments of extreme statistics.
{\it Annals of Mathematical Statistics}, {\bf 64}, 1738--1745.


\bibitem{}
Novak, S. Y. and Utev, S. A. (1990).
Asymtotics of the distributio of the ratio of sums of random variables.
{\it Siberian Mathematical Journal}, {\bf 31}, 781--788.


\bibitem{}
Pickands, J. (1968).
Moment convergence of sample extremes.
{\it Annals of Mathematical Statistics}, {\bf 39}, 881--889.


\bibitem{}
Polfeldt, T. (1970).
The order of the minimum variance in a non-regular case.
{\it Annals of Mathematical Statistics}, {\bf 41}, 667--672.


\bibitem{}
Ramachandran, G. (1984).
Approximate values for the moments of extreme order statistics in large samples. Statistical extremes and applications (Vimeiro, 1983), pp. 563--578,
NATO Advanced Science Institutes Series C: Mathematical and Physical Sciences, 131, Reidel, Dordrecht.


\bibitem{}
Resnick, S. I. (1987).
{\it Extreme Values, Regular Variation, and Point Processes}.
Springer--Verlag, New York.


\bibitem{}
Stuart, A. and Ord, J. K. (1987).
{\it Kendall's Advanced Theory of Statistics}, 5th edition, Volume 1.
Griffin, London.


\bibitem{}
Withers, C. S. and Nadarajah, S. (2007a).
Asymptotic multivariate distributions and moments of extremes.
{\it Technical Report}, Applied Mathematics Group, Industrial Research Ltd., Lower Hutt, New Zealand.


\bibitem{}
Withers, C. S. and Nadarajah, S. (2007b).
Expansions for the beta function and its inverse when one parameter is large.
{\it Technical Report}, Applied Mathematics Group, Industrial Research Ltd., Lower Hutt, New Zealand.


\end{thebibliography}
\end{document}